# Structures of Molecules at the Atomic Level: Caffeine and Related Compounds

Raji Heyrovska (1) and Saraswathi Narayan (2)

((1) Academy of Sciences of the Czech Republic: Email: rheyrovs@hotmail.com, (2) Villa Julie College, Stevenson, MD, Email: f-naraya@mail.vjc.edu)

**Abstract**

Recent rsearches have shown that the lengths of the chemical bonds, whether completely or partially covalent or ionic, are sums of the radii of the adjacent atoms and/or ions. On investigating the bond length data for the molecular components of nucleic acids, all were found (for the first time) to be effectively the sums of the covalent radii of the adjacent atoms. This work shows that the bond lengths in caffeine and related molecules are likewise sums of the covalent radii of C, N, O and H. This has enabled arriving at the atomic structures of these molecules, also for the first time.

*(This work was presented at the 10th Eurasia Conference on Chemical Sciences, Manila, Philippines, January 2008.)*

## 1. Introduction

It was shown in recent years [1-3] that bond lengths of completely or partially covalent or ionic bonds are sums of the radii of the adjacent atoms or ions. The atomic radii are the covalent radii [4] or the bonding atomic radii [5] d(A) defined as,

$$d(A) = d(AA)/2 \qquad\qquad\qquad\qquad (1)$$

where $d(AA)$ is the interatomic distance. The ionic radii are the Golden sections of $d(AA)$, based on the Golden ratio $\phi$, [1,2]. The Golden ratio ($\phi = 1.618$), also known as The Divine ratio, appears in the geometry of a variety of Nature's creations, as described e.g. in [6]. It is the ratio a/b of a larger segment (a) to a smaller segment (b) of a line which are such that a/b = (a + b)/a. From the latter, one obtains the Golden quadratic equation, $a/b = (a/b)^2 - 1$ whose positive root is $(a/b) = (1 + 5^{1/2})/2 = \phi$ (= 1.618..). The segements $a = c/\phi$ and $b = c/\phi^2$ are called the Golden sections of the line of length $c = a + b$. Note that $(1/\phi) + (1/\phi^2) = 1$. A simple way of visualizing $\phi$ is that it is eaqul to the ratio of the diagonal to a side in a regular penatgon.

It was shown [1] (for the first time) that, in fact, the Golden ratio arises right in the core of the atom, as follows: the ground state Bohr radius for H ($a_B = 0.53$ Å) is related to the ionization potential, $I_H = (e/2\kappa a_B)$, which in turn is the sum (or the absolute difference in potentials) of $I_p$ ($= e/2\kappa a_p$) and $I_e$ ($= -e/2\kappa a_e$) of $p^+$ and $e^-$, respectively, where $a_B = a_p + a_e$ and $\kappa = 4\pi\varepsilon$ is the electrical permittivity. From these, one obtains the following relations:

$$a_B = a_p + a_e; \ 1/a_B = (1/a_p) - (1/a_e) \qquad\qquad\qquad (2a,b)$$

$$(a_e/a_p)^2 - (a_e/a_p) - 1 = 0 \quad \text{(Golden quadratic, obtained from Eqs. 2a,b)} \qquad (3a)$$

$$(a_e/a_p) = \phi = (1 + 5^{1/2})/2 = 1.618.. \text{ (positive root)} \qquad\qquad (3b)$$

where $a_p = (a_B/\phi^2)$ and $a_e = (a_B/\phi)$ are the <u>Golden sections</u> of $a_B$.

A square with $a_B$ as a side has two protons and two electrons at opposite corners. The diagonal of this square gives [1a] the covalent bond length, $d(HH) = 0.74$ Å of the $H_2$ molecule [4]. The covalent radius $d(H) = d(HH)/2 = 0.37$ Å. Since $a_B$ has two Golden sections, one gets the following relations:

$$d(HH) = 2d(H) = 2^{1/2}a_B = 2^{1/2}(a_e + a_p) = d(H^-) + d(H^+) \qquad (4a)$$

$$d(H^+) = d(HH)/\phi^2 \text{ and } d(H^-) = d(HH)/\phi \qquad (4b,c)$$

where $d(H^-) = 2^{1/2}a_e = 0.46$ Å and $d(H^+) = 2^{1/2}a_p = 0.28$ Å are the anionic and cationic radii of H, respectively, and are the Golden sections of $d(HH)$. Note also that $2^{1/2}d(H) = a_B$ is the diagonal of a square with $d(H)$ as a side. $H^-$ and $H^+$ correspond [1] to the ionic structures in the resonance structures suggested by Pauling [4] for the $H_2$ molecule.

A survey of the literature shows that $d(H^+) = 0.74/\phi^2 = 0.28$ Å is actually the empirical radius for hydrogen suggested by Pauling [4] for explaining the partially ionic bonds in hydrogen halides (HX), where X has the covalent radius. It can be seen therefore that the partial ionic charcater of the HX bond is due to $d(H^+)$.

It was found [1] that on subtracting $d(H^+) = 0.28$ Å, from the known [4] bond lengths d(HX) of hydrogen halides and d(MH) of alklai metal hydrides, where X = F, .., I and M = Li, .., Cs, one obtains d(X) and $d(M^+)$, respectively, as shown in Eqs. 5a,b. d(X) is found to be half the covalent bond distance d(XX) and $d(M^+)$ is a Golden section of the crystallographic lattice distance, d(MM) as shown:

$$d(HX) - d(H^+) = d(XX)/2 = d(X); \quad \text{(for HX)} \qquad (5a)$$

$$d(MH) - d(H^+) = d(MM)/\phi^2 = d(M^+); \quad \text{(for MH)} \qquad (5b)$$

On subtracting the values of $d(M^+)$ obtained from Eq. 5b from the known [4] interionic distances $d(MX)$ in alkali halides, one obtains $d(X^-)$, which is a Golden section of $d(XX)$ as shown [1]:

$$d(MX) - d(M^+) = d(XX)/\phi = d(X^-); \quad \text{(for MX)} \tag{5c}$$

These show that the covalent atomic radii and Golden-ratio based ionic radii are additive, and that like $d(HH)$, both $d(MM)$ and $d(XX)$ have two Golden sections which are the anionic and cationic radii of M and X .

For covalent bonds such as CH, NH, OH, etc., [1-4], the bond length is the sum of covalent radii $d(A) = d(AA)/2$ and $d(B) = d(BB)/2$ as given by,

$$d(AB) = d(A) + d(B) \tag{6}$$

Thus, many bond lengths were shown [1] to be sums of the radii of the two atoms/ions constituting the bonds. The additivity of atomic/ionic radii was found to hold even in aqueous solutions [7] and also for hydrogen bonds [8] in many inorganic and biochemical groups including in the Watson-Crick base pairs in DNA and RNA.

## 2. Completely covalent bonds in DNA and RNA [3]

On investigating the bond length data for biological molecules like the molecular componenets of DNA and RNA, it was found [3] that all the skeletal bond lengths are sums of the appropriate covalent radii of the adjacent atoms, C, N, O, H and P. These covalent radii are shown in Fig. 1a. Fig. 1b shows how these atomic radii fit the 34 : 20 A section of the DNA (or RNA) helix. The details can be found in [3].

### 3. Completely covalent bonds in caffeine and related molecules:

Proceeding next to another set of improtant biological molecules, namely, caffeine [9], its liver metabolites and xanthine [10], it is shown here (for the first time) that the atomic radii in Fig. 1a also explain all the bond lengths. Table 1a shows the average bond lengths (+/- 0.03 Å) in the literature [11-13] and the corresponding sums of atomic covalent radii, R(sum) using the data in Fig. 1a. In Tab. 1b, can be seen a comparison of the radii sum R(sum) with the overall average of the bond lengths [11-13] from Tab. 1a.

A graph of the average bond lengths [11-13 ] (data in Tabs. 1a,b) versus R(sum) is shown in Fig. 2. The average values fall on a least square straight line with slope = 1.02 and intercept, -0.04. It can be seen that the values of R(sum), calc. in the last column in Tab. 1b (using the slope and intercept of the least square line) agree well with R(sum) in col. 3. Thus all the bonds in caffeine and the related molecules are covalent and the bond lengths can be considered as sums of the atomic covalent radii.

Thus the conventional molecular structures (see Fig.3, left top [10]) of these compounds have been transformed here into the "atomic structures" as shown in Fig. 3, where all the bond lengths are sums of the appropriate covalent radii of he adjacent atoms.

### Acknowledgements

R. H. is grateful for support by grants AVOZ50040507 of the Academy of Sciences of the Czech Republic and LC06035 of the Ministry of Education, Youth and Sports of the Czech Republic. R. H. also thanks the Organizers  of the 10th Eurasia Conference on Chemical Sciences, Jan. 2008, Manila, Philippines, for the financial support to participate in this conference. S. N. thanks VJC for their partial financial support to attend this conference.

**References**

[1] Heyrovska R., <u>Mol. Phys</u>. 2005; 103: 877-882.

[2] Heyrovska R., in: <u>Innovations in Chemical Biology</u>, Proceedings of the 9th Eurasia Conference on Chemical Sciences, Antalya, Turkey, 2006, (Springer.com, ISBN-978-1-4020-6954-3, expected publication date: May 2008), Ed: Bilge Sener.

[3] Heyrovska R., a) <u>arXiv:0708.1271v4</u>; b) <u>Open Structural Biology Journal</u>, 2008; 2, 1-7.

[4] Pauling L., <u>The Nature of the Chemical Bond</u> (Cornell Univ. Press, NY, 1960).

[5] <u>http://wps.prenhall.com/wps/media/objects/3311/3390919/blb0702.html</u>

[6] <u>http://www.goldennumber.net/</u> (and the literature therein).

[7] Heyrovska R., <u>Chem. Phys. Lett</u>. 2006; 429: 600-605, <u>doi:10.1016/j.cplett.2006.08.073</u>

[8] Heyrovska R., <u>Chem. Phys. Lett</u>. 2006; 432: 348-351, doi:10.1016/j.cplett.2006.10.037

[9] Zajec M.A., Zakrzewski A.G., Kowal M.G., Narayan S., <u>Synthetic Communs.</u>, 2003; 33: 3291-97.

[10] (<u>http://en.wikipedia.org/wiki/Caffeine</u>)

[11] Ucun F., Saglam A., Guclu V., <u>Spectrochim. Acta Part A</u>, 2007; 67: 342-349.

[12] Egawa T., Kamiya A., Takeuchi H., Konaka S., <u>J. Mol. Struc</u>., 2006; 825: 151-157

[13] Gunasekaran S., Sankari G., Ponnusamy S., <u>Spectrochim. Acta Part A</u>, 2005; 61: 117-127.

[14]: Franklin R.E., Gosling R.G., <u>Nature</u>, 1953; 171: 740-741; see for full texts of this and other papers: <u>http://www.nature.com/nature/dna50/archive.html</u>

**Table 1a**.  Average bond lengths (+/- 0.03 Å) from [11-13] and the corresponding sums of atomic covalent radii, R(sum). See Fig. 1a for the radii of the atoms which are in the atomic structures in Fig. 3, which also shows the numbering of the atoms.

| 1 | 2 | 3 | 4 | 5 | 6 | 7 | 8 | 9 | 10 | 11 | 12 |
|---|---|---|---|---|---|---|---|---|---|---|---|
| | | <------------------------------------#Average bond lengths -------------------------------------------> | | | | | | | | | |
| Bonds | $^{\$}$R(sum) | Caf. [11]theor | Theobr. [11]theor | Xan. [11]exp | Xan. [11]theor | Caf. [12]exp | Caf. [12]theor | Xan. [13]exp | Caf. [13]exp | Theophy. [13]exp | Theobr. [13]exp |
| $C_{d.b}$ - $O_{d.b}$ | 1.27 | 1.21 | 1.21 | 1.23 | 1.21 | 1.21 | 1.23 | 1.22 | 1.23 | 1.21 | 1.21 |
| $C_{d.b}$ - $N_{d.b}$ | 1.29 | 1.34 | 1.34 | 1.35 | 1.34 | 1.33 | 1.35 | 1.32 | 1.33 | 1.32 | 1.32 |
| $C_{d.b}$ - $N_{s.b}$ | 1.37 | 1.38 | 1.38 | 1.37 | 1.37 | 1.38 | 1.38 | 1.35 | 1.37 | 1.36 | 1.36 |
| $C_{res}$ - $C_{d.b}$ | 1.39 | 1.40 | 1.40 | 1.39 | 1.40 | 1.40 | 1.40 | 1.42 | 1.38 | 1.39 | 1.41 |
| $C_{res}$ - $N_{s.b}$ | 1.42 | 1.39 | 1.38 | 1.38 | 1.38 | 1.38 | 1.38 | 1.34 | 1.41 | 1.34 | 1.34 |
| $C_{s.b}$ - $N_{s.b}$ | 1.47 | 1.46 | 1.46 | | | 1.46 | 1.46 | | 1.48 | 1.47 | 1.47 |
| $C_{d.b}$ – H | 1.04 | 1.08 | 1.08 | | 1.08 | 1.09 | 1.08 | 0.98 | 1.15 | 0.98 | 0.98 |
| $N_{s.b}$ – H | 1.07 | | 1.00 | | 1.00 | | | 0.98 | | 0.98 | 0.98 |
| $C_{s.b}$ – H | 1.14 | 1.08 | 1.09 | | | 1.09 | 1.09 | | | | |

**Footnotes:**
#**Abbreviations**: **Caf.**: Caffeine, **Theobr.**: Theobromine, **Theophy.**: Theophylline, **Xan.**: Xanthine.
$^{\$}$**Covalent radii (Å)**: N(1) = N(3) = N(7) = $N_{s.b.}$ = **0.70**; N(9) = $N_{d.b.}$ = **0.62**; C(2) = C4 = C(6) = C(8) = $C_{d.b.}$ = **0.67**; C(5) = $C_{res}$ = **0.72**; C(1) = C(3) = C(7) = $C_{s.b.}$ = **0.77**; O(2) = O(6) = $O_{d.b.}$ = **0.60**; all **H = 0.37.**

**Table 1b**. Comparison of the sum of atomic radii R(sum) (col. 3) with the overall average (col. 4) of the bond lengths [11-13] in cols. 3 -12, Tab. 1a. See Fig.3 for the numbering (col. 1) and Fig. 1a for the type (col. 2) of the atoms. Note that R(sum), calc. (l.sq.line in Fig. 2) in col. 5 reproduces R(sum) in col. 3.

| 1 | 2 | 3 | 4 | 5 |
|---|---|---|---|---|
| Bonds of the same length | Bonds | R(sum) | Average [7 – 9] | R(sum), calc.: l.sq.line |
| C2 – O2 = C6 - O6 = | $C_{d.b}$ - $O_{d.b}$ | 1.27 | 1.22 | 1.26 |
| C4 – N9 = N9 - C8 = | $C_{d.b}$ - $N_{d.b}$ | 1.29 | 1.33 | 1.28 |
| C2 - N1 = N1 - C6 = C2 – N3 = N3 – C4 = | $C_{d.b}$ - $N_{s.b}$ | 1.37 | 1.37 | 1.36 |
| C4 – C5 = C5 - C6 = | $C_{res}$ - $C_{d.b}$ | 1.39 | 1.40 | 1.38 |
| C5 – N7 = | $C_{res}$ - $N_{s.b}$ | 1.42 | 1.37 | 1.41 |
| N1 – C1 = N3 - C3 = N7 – C7 = | $C_{s.b}$ - $N_{s.b}$ | 1.47 | 1.46 | 1.46 |
| C8 – H8 = | $C_{d.b}$ – H | 1.04 | 1.09 | 1.02 |
| N1 – H1 = N3 - H3 = N7 – H7 = | $N_{s.b}$ – H | 1.07 | 0.99 | 1.05 |
| C1 - H1, H1', H1" = C3 – H3, H3', H3" = C7 – H7, H7', H7" = | $C_{s.b}$ – H | 1.14 | 1.09 | 1.12 |

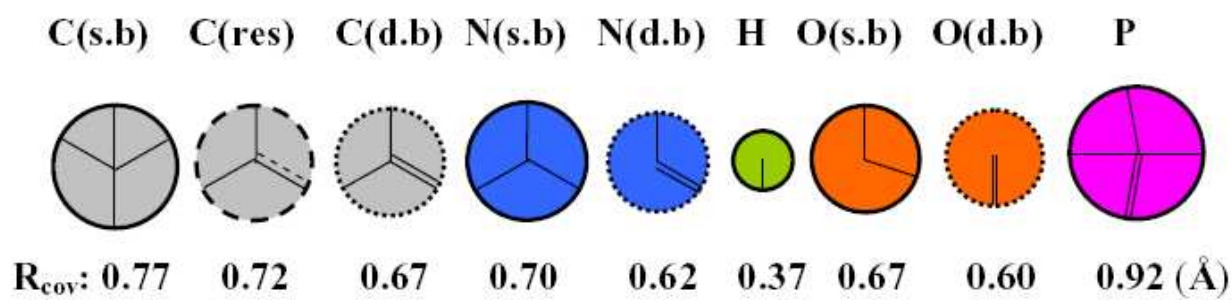

**Fig. 1a.** Covalent radii [4] of the atoms: C, N, H, O & P which account [3] for all the bond lengths in DNA and RNA. Subscripts: s.b: single bond (C: as in diamond), res.: resonance bond (C: as in graphite and benzene), d.b: double bond.

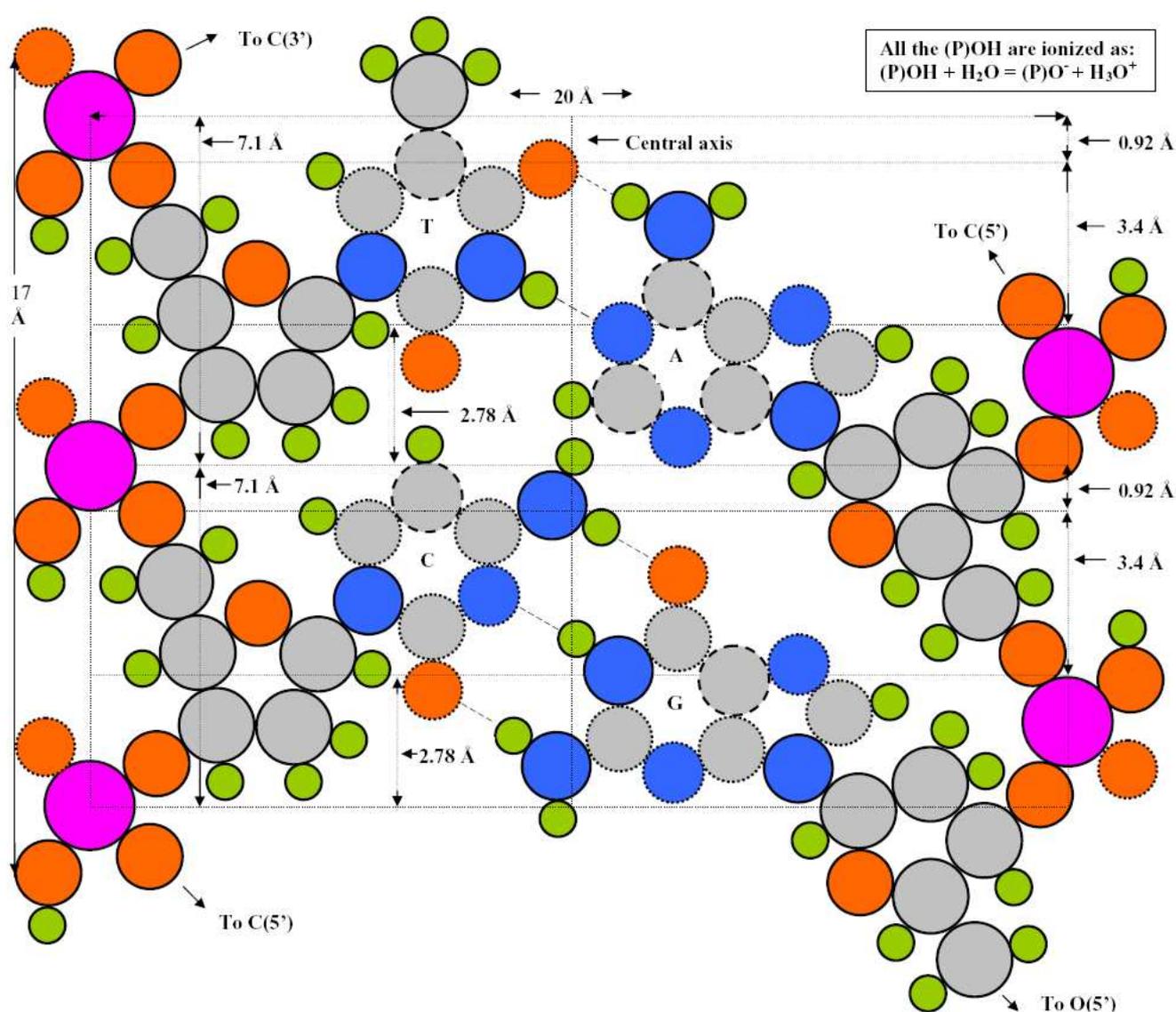

**Fig. 1b**. Atomic structures of the molecules in DNA nucleotides [3], with bond lengths = sums of covalent radii of adjacent atoms. In the 17 Å section, there are 5P atoms (this is half the 34 : 20 Å section per turn of the helix with ten P atoms, and each P atom is 10 Å from the central axis [14].) RNA has U and ribose in place of T and deoxyribose. The lengths of the hydrogen bonds in T & A and C & G are explained in [8].

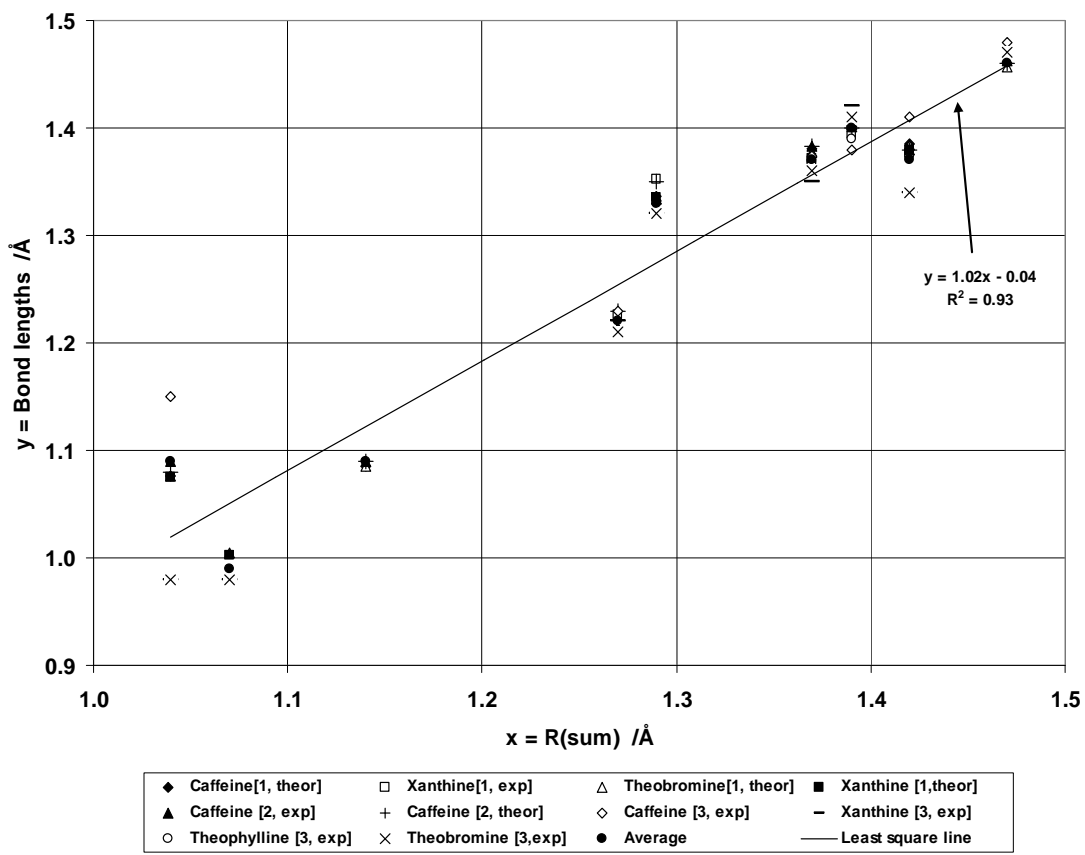

**Fig. 2.** Graph of the average bond lengths in caffeine and related molecules [11-13 ] (data in Tabs. 1a,b) versus R(sum). The values of R(sum), calc. in Tab. 1b (using the least square slope, 1.02 and intercept, -0.04) agree well with R(sum) in col. 3.

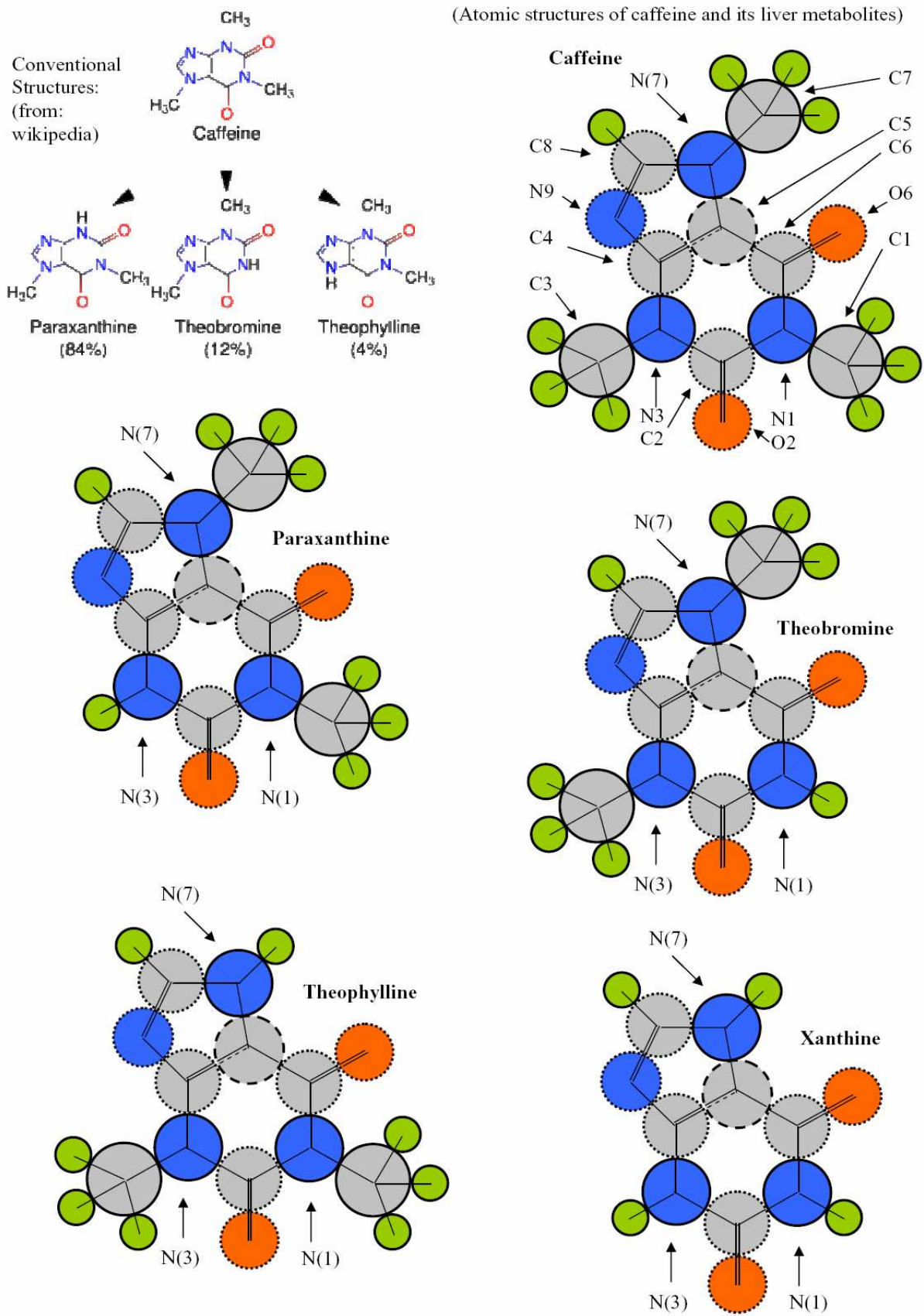

**Fig. 3**. "Atomic structures" of caffeine and related molecules: Bond lengths = sums of atomic covalent radii (see Fig. 1a). Left top: conventional structures from [10].